\pdfoutput=1
\documentclass[%
reprint,
amsmath,amssymb,
aps,
pra,
]{revtex4-1}

\usepackage{graphicx,wrapfig,lipsum}
\usepackage{subeqnarray}
\usepackage{color}

\usepackage{natbib}

\newcommand{\req}[1]{Eq.\,({\ref{#1}})}

\begin{document}
\title{Classical neutral point particle in linearly polarized EM plane wave field}
\author{Martin Formanek}
\email{martinformanek@email.arizona.edu}
\author{Andrew Steinmetz}
\author{Johann Rafelski}
\affiliation{Department of Physics\\
University of Arizona\\
Tucson, AZ, 85719, USA}

\date{\today}
%
\begin{abstract}
We study a covariant classical model of neutral point particles with magnetic moment interacting with external electromagnetic fields. Classical dynamical equations which reproduce a correct behavior in the non-relativistic limit are introduced. We also discuss the non-uniqueness of the covariant torque equation. The focus of this work is on Dirac neutrino beam control. We present a full analytical solution of the dynamical equations for a neutral point particle motion in the presence of an external linearly polarized EM plane wave (laser) fields.  Neutrino beam control using extremely intense laser fields could possibly demonstrate Dirac nature of the neutrino. However, for linearly polarized ideal laser waves we show cancellation of all leading beam control effects.
\begin{description}
\item[PACS numbers]
21.10.Ky Electromagnetic moments, 03.30.+p Special relativity, 13.40.Em Electric and magnetic moments 
\end{description}
\end{abstract}
\maketitle


\section{Introduction}
Since neutral particles (such as neutrons or Dirac neutrinos) posses a magnetic moment they experience a Stern-Gerlach type force when exposed to external fields. We invoke a theoretical framework developed for charged particles in our previous paper \cite{rafelski2018relativistic}.

In a first look on the environment of the neutral particle -- laser interactions using our framework \cite{formanek2017strong} we focused on describing the acceleration force neutral particles experience. We found out that the square root of invariant acceleration is greatly enhanced by the relativistic factor $\gamma_0^2$ with which the particle enters the field. This motivated our present study in which we aim to describe the motion of the neutral particles by solving the dynamical equations.   

Classical solution for the charged particle dynamics in a linearly polarized external plane wave field is well known \cite{sarachik1970classical,itzykson2005quantum, rafelski2017electrons}. The objective of our present work is to expand on this knowledge by describing how this EM field configuration affects neutral particles with magnetic moment. In doing so we employ mathematical methods developed in context of solution of the Landau-Lifshitz equation for charged particles in an external plane wave field \cite{hadad2010effects,di2008exact}. Specifically we adapt technique of projecting the 4-velocity and spin 4-vector on the wave vector and polarization vector of the field and solving differential equations for these projections.

As our application we focus on laser interactions with ultra-relativistic neutrinos. Neutrinos remain the least understood elementary particles \cite{mohapatra2007theory} and there is current intense interest in furthering comprehension of their elementary properties. Perhaps the most elementary  question about  the neutrino is if it is a Dirac or Majorana particle. Today, there is intense interest in search for double beta decay \cite{kotani1985double} with new experiments planned \cite{henning2016current} which could prove that neutrinos are Majorana particles.

On the other hand we cannot dismiss possibility that neutrinos are Dirac particles. A minimal extension of the Standard model \cite{fujikawa1980magnetic} gives us a lower bound for the magnetic moment of the Dirac neutrino
\begin{equation}\label{eq:lower}
\mu_i \approx 3.2 \times 10^{-19} \left(\frac{m_i}{eV} \right) \mu_B\;,
\end{equation}
where $\mu_B = e\hbar/2m_e$ is the Bohr magneton. The $m_i$ is the neutrino mass eigenstate $\nu_i$, which for electron neutrino has scale $m_\nu \leq 0.2$ eV.  On the other hand neutrino magnetic moment cannot be too large considering variety of neutrino driven processes in astrophysics. Present experimental upper bound for the magnetic moment is \cite{patrignani2016review}
\begin{equation}\label{eq:upper}
\mu_\nu < 2.9 \times 10^{-11} \mu_B\;.
\end{equation}
Detection of neutrino magnetic moment in this range would be  complementary to a (so far nill)
result of the search for double beta decay since it is believed that a Majorana neutrino just like a photon cannot have a magnetic moment. Any experiment exploiting and detecting neutrino  magnetic moment would therefore proof their Dirac nature.  

This study was motivated by the possibility of exploring control of neutrino beams by ultra-laser intense laser fields. Having such capabilities would be of an utmost importance because manipulating neutrino beams through a well defined interaction could resolve the question of Dirac/Majorana character of neutrinos and could provide an opportunity to directly measure their mass and magnetic moment.

 At the present moment, there is a multitude of competing models which approach relativistic dynamics of a point particle from first principles. Thomas and Frenkel  \cite{thomas1926motion,frenkel1926elektrodynamik} introduced in 1926 Frenkel's equation of motion for spin as a second rank tensor. Bargmann, Michel, and Telegdi formulated in late 1950s TBMT equations   \cite{bargmann1959precession} which are until today often used for classical description of the spin dynamics despite missing the Stern-Gerlach-like force. Another formulation is looking on a classical limit of the relativistic quantum-mechanical Dirac equation using the Foldy-Wouthuysen transformation \cite{silenko2008foldy}. Unfortunately, the generalization of relativistic quantum description for particles with anomalous magnetic moment is still not clear \cite{steinmetz2019magnetic}. An overview of these approaches and their numerical tests can be found in \cite{wen2016identifying,wen2017spin}. Our model \cite{rafelski2018relativistic} expands on the pioneering work of the first formulations. We incorporate force on a magnetic dipole to the TBMT equations with an arbitrary anomalous magnetic moment.

The interaction between neutrinos and external plane wave field can be also treated within bounds of quantum field theory. This has been shown in \cite{skobelev1991interaction} for a specific model of anomalous magnetic moment/electric dipole interaction to describe decay $\gamma \rightarrow \nu \bar{\nu}$, and bremsstrahlung processes. We are interested in the classical behavior, because except for the relic neutrinos, the neutrinos which we typically encounter (from nuclear reactions in the Sun, our nuclear reactors, radioactive decay in Earth or sources outside of are solar system) are ultra-relativistic and within the classical limit for visible laser light
\begin{equation}\label{eq:classlim}
\frac{\lambda_\nu}{\lambda_\gamma} \ll 1\;,
\end{equation}
where $\lambda_\nu$ and $\lambda_\gamma$ are de Broglie wavelength of a neutrino with a given energy and wavelength of the light respectively.

Our paper is organized in three sections. In section \ref{sec:dynamics} we present the dynamical equations for the neutral particles and provide a rationale why we have chosen their particular form. In the section \ref{sec:solution} we study a specific case of external linearly polarized plane wave field. In section \ref{sec:neutrinos} we discuss the case of ultra-relativistic neutrinos.  

Notation remark:  we use the following convention for the metric tensor $g_{\mu\nu}$ and totally anti-symmetric covariant pseudo-tensor $\epsilon_{\mu\nu\alpha\beta}$
\begin{equation}
\text{diag}\; g_{\mu\nu} = \{1,-1,-1,-1\}, \quad \epsilon_{0123} = - \epsilon^{0123} = +1\;,
\end{equation}
also SI units are used throughout.

\section{Formulation of neutral particle dynamics}\label{sec:dynamics}
In the paper \cite{rafelski2018relativistic} we introduced a generalized Lorentz Force equation which allows us to account for a Stern-Gerlach force acting on a magnetic moment of point particle in the presence of external electromagnetic fields
\begin{equation}\label{eq:udynam}
\dot{u}^\mu = \frac{1}{m} (e F^{\mu \nu} - d s \cdot \partial F^{*\mu\nu})u_\nu\;,
\end{equation}
where $d$ is the constant of proportionality between particle spin and magnetic moment
\begin{equation}\label{eq:magmoment}
|\pmb{\mu}| = c d |\pmb{s}|, \quad \pmb{\mu} = \mu \frac{\pmb{s}}{|\pmb{s}|}\;,
\end{equation}
superscript $*$ denotes a dual tensor 
\begin{equation}
F^{*\mu\nu} = \frac{1}{2}\epsilon^{\mu\nu\alpha\beta}F_{\alpha\beta}\;,
\end{equation}
and a \lq dot\rq\ means a derivative with respect to proper time $\tau$. As was discussed in \cite{rafelski2018relativistic} the form of the corresponding dynamical torque equation for spin is not unique. Similar non-uniqueness manifests itself also in the quantum case \cite{steinmetz2019magnetic} where extensions to Dirac and Klein-Gordon Pauli equations to accommodate magnetic moment differ substantially. In the classical case we can only demand that the spin dynamics is consistent with the particle motion dynamics
\begin{equation}\label{eq:constraint1}
u \cdot s = 0\;,
\end{equation}
and in the instantaneous frame co-moving with the particle the equation for the spin dynamics has to contain a torque term
\begin{equation}\label{eq:constraint2}
\frac{d\pmb{s}}{dt} = \pmb{\mu} \times \pmb{\mathcal{B}}\;,
\end{equation}
which ensures the correct torque behavior when magnetic moment of particle is trying to align itself with the external field \cite{morley2015instantaneous}. This observation discussed before for charged particles has to be true for a neutral particle as well. 

We presented a viable choice for the spin dynamics satisfying both constraints \cite{formanek2017strong} as will be also shown explicitly in section \ref{sec:nonrel}
\begin{multline}\label{eq:spindynam}
\dot{s}^\mu = \frac{e}{m}F^{\mu\nu}s_\nu + \widetilde{a} \left( F^{\mu\nu}s_\nu - \frac{u^\mu}{c^2} u \cdot F \cdot s\right)\\ 
- \frac{1}{mc}\left(\frac{e}{m} + \widetilde{a}\right) s \cdot \partial F^{*\mu\nu}s_\nu\;.
\end{multline}
The first term ensures consistency with the Lorentz force, second term introduces a magnetic anomaly for the charged particles $\widetilde{a}$ satisfying
\begin{equation}
dc = \frac{e}{m} + \widetilde{a}
\end{equation}
and the third term is necessary for consistency with the Stern-Gerlach term in \req{eq:udynam} through derivative of the condition \req{eq:constraint1}. We have chosen this form of writing the constants because we can easily perform the limit of neutral particles, when charge $e = 0$ and the whole magnetic moment is anomalous \cite{sakurai1967advanced}. This allows us to write for neutral particles a following set of equations
\begin{align}
\label{eq:firstdynam} 
\dot{u}^\mu &=  - s\cdot\partial F^{*\mu\nu}u_\nu \frac{d}{m}\;,\\
\label{eq:seconddynam} 
\dot{s}^\mu &= cd \left(F^{\mu\nu}s_\nu - \frac{u^\mu}{c^2}(u\cdot F \cdot s)\right)- s\cdot\partial F^{*\mu\nu}s_\nu \frac{d}{m}\;.
\end{align}
These dynamical equations for neutral particles will be a starting point of our study. 
\subsection{Justification: non-relativistic behavior}\label{sec:nonrel}
In the laboratory frame the velocity and spin and spin 4-vectors read
\begin{equation}
u^\mu = \gamma c (1, \pmb{\beta}), \qquad s^\mu = (\pmb{\beta} \cdot \pmb{s}, \pmb{s})\;,
\end{equation}
where the spin 3 vector is a Lorentz transformation of the instantaneous co-moving frame spin $\pmb{s}_c$
\begin{equation}\label{eq:dsdt}
\pmb{s} = \pmb{s}_c + \frac{\gamma - 1}{\beta^2} (\pmb{\beta} \cdot \pmb{s}_c)\pmb{\beta}\;.
\end{equation}
The electromagnetic tensor and its dual in the laboratory frame are
\begin{equation}
F^{\mu\nu} = \left(
\begin{matrix}
0 &-\pmb{E}/c\\
\pmb{E}/c & -\epsilon_{ijk} B^k\\
\end{matrix} \right), \quad F^{*\mu\nu} = \left(
\begin{matrix}
0 & \pmb{B}\\
-\pmb{B} & \epsilon_{ijk}E^k/c
\end{matrix}\right)\;.
\end{equation}
Finally, the covariant gradient term can expressed as
\begin{equation}
s \cdot \partial = \pmb{\beta} \cdot \pmb{s}\frac{\partial}{\partial c t} + \pmb{s} \cdot \nabla 
\end{equation}
The spatial part of the force equation \req{eq:firstdynam} reads
\begin{equation}\label{eq:nonrel1}
\frac{d}{dt}(\gamma \pmb{\beta}) = \frac{d}{m}\left(\pmb{\beta} \cdot \pmb{s}\frac{\partial}{\partial c t} + \pmb{s} \cdot \nabla \right)(\pmb{B} c - \pmb{\beta} \times \pmb{E})\;,
\end{equation}
and the spatial part of the torque equation \req{eq:seconddynam} is
\begin{multline}\label{eq:nonrel2}
\gamma \frac{d}{dt}\pmb{s} = d \left(\pmb{E}(\pmb{\beta} \cdot \pmb{s}) + \pmb{s}\times\pmb{B}c \right)\\
-d\gamma^2 \pmb{\beta} (\pmb{E} \cdot \pmb{s} - (\pmb{\beta} \cdot \pmb{E})(\pmb{\beta} \cdot \pmb{s}) - \pmb{\beta} \cdot (\pmb{s} \times \pmb{B}c)\\
- \frac{d}{mc}(s \cdot \partial)(\pmb{s} \times \pmb{E} - \pmb{B}c(\pmb{\beta} \cdot \pmb{s}))
\end{multline}
Following the derivation presented by Schwinger \cite{schwinger1974spin} we can substitute \req{eq:nonrel1} into \req{eq:nonrel2}  and in the non-relativistic limit we neglect terms quadratic in $\pmb{\beta}$ and higher
\begin{multline}
\dot{\pmb{s}} \approx d \left(\pmb{E}(\pmb{\beta} \cdot \pmb{s}) - \pmb{\beta}(\pmb{E} \cdot \pmb{s}) + \pmb{s}\times\pmb{B}c \right)\\
+ \dot{\pmb{\beta}}(\pmb{\beta} \cdot \pmb{s}) - \frac{d}{mc}(\pmb{s} \cdot \nabla)(\pmb{s}\times \pmb{E})
\end{multline}
Substituting into the left hand side derivative of \req{eq:dsdt} and combining with the term containing $\dot{\pmb{\beta}}$ on the right hand side we obtain the Thomas precession term
\begin{eqnarray}\label{eq:thomas}
\left(\frac{d}{dt}\pmb{s}_c\right)_\text{TP} \approx \frac{1}{2}\pmb{\beta} \times \dot{\pmb{\beta}} \times \pmb{s}_c\;,
\end{eqnarray}
which is independent of the size of the magnetic moment.

In the instantaneous frame co-moving with the particle we have $\pmb{\beta} = 0$, $\gamma = 1$ which further simplifies our equations \req{eq:nonrel1},\ref{eq:nonrel2} to
\begin{align}
\label{eq:force} \frac{d}{dt}(\pmb{v})|_\text{c} = &= \frac{1}{m}(\pmb{\mu} \cdot \pmb{\nabla}) \pmb{\mathcal{B}}\;, \\
\label{eq:torque}\frac{d}{dt}\pmb{s}|_\text{c} & = \pmb{\mu} \times \left(\pmb{\mathcal{B}} - \frac{1}{mc} (\pmb{s} \cdot \pmb{\nabla}) \frac{\pmb{\mathcal{E}}}{c}\right) \;,
\end{align}
where we also rewrote spins in terms of the magnetic moment using relationship \req{eq:magmoment}. These expressions have all the desired properties. The force equation \req{eq:force} contains Gilbertian form of Stern-Gerlach interaction \cite{rafelski2018relativistic}. The torque equation \req{eq:torque} behaves according to constraint in \req{eq:constraint2} - spin aligning with the magnetic field. In addition we have another term which depends on the component-wise gradient of the electric field in the direction of the spin. This term is a new prediction of our theory, but a necessary addition in order to make the torque equation \req{eq:seconddynam} compatible with the Stern-Gerlach force equation \req{eq:firstdynam} through constraint \req{eq:constraint1} which is not accounted for in the standard TBMT formulation \cite{bargmann1959precession}.  A similar term, depending on quantum mechanical model of spin, also arises from a classical limit of relativistic quantum equations, we will explore this correspondence under a separate cover. Such additional force causes the spin vector not only to align with the magnetic field, but also depend on the gradient of the electric field.
\subsection{Non-uniqueness of the torque equation}\label{sec:nonuni}
The proposed torque equation for the neutral particle \req{eq:torque} is the simplest form which is consistent with the equation for the force \req{eq:force} and generates the correct torque term in the frame co-moving with the particle. We could imagine other terms, orthogonal to $u^\mu$ which could be included in the torque equation. For example we can add
\begin{equation}
\dot{s}^\mu = \ldots + \widetilde{b} \left(s \cdot \partial F^{*\mu\nu}s_\nu - \frac{u^\mu}{c^2}u \cdot (s \cdot \partial)F^* \cdot s \right)\;,
\end{equation}
with $\widetilde{b}$ being another constant characterizing the classical point particle. If we would repeat the analysis in the section \ref{sec:nonrel} with this addition, the Thomas precession term \req{eq:thomas} would be sensitive to the value of $\widetilde{b}$. A precession experiment with neutral particles could resolve if such modification is necessary.  

\section{Solution for linearly polarized plane wave}\label{sec:solution}
We consider potential of  a planelinear polarized electromagnetic  wave 
\begin{equation}\label{eq:plane wave}
A^\mu = \varepsilon^\mu \mathcal{A}_0 f(\xi), \quad \xi = \frac{\omega}{c}\hat{k} \cdot x\;,
\end{equation}
where $\varepsilon^\mu$ is polarization of the plane wave; $\xi$ its invariant phase; $\mathcal{A}_0$ amplitude; and $\hat{k}^\mu$ a unit-less vector in the direction of the wave vector. The wave vector is time-like and transverse to the space-like  polarization vector
\begin{equation}\label{eq:properties}
\quad \hat{k}^2 = 0, \quad \hat{k} \cdot \varepsilon = 0, \quad  \varepsilon^2 = -1\;. 
\end{equation}
 $f(\xi)$ is a function characterizing the laser pulse containing both the oscillatory part, and the pulse envelope. 

Using this 4-potential we can construct an electromagnetic field tensor
\begin{equation}\label{eq:emtensor}
F^{\mu\nu} = \partial^\mu A^\nu - \partial^\nu A^\mu = \frac{\mathcal{A}_0 \omega}{c}f'(\xi) (\hat{k}^\mu \varepsilon^\nu - \varepsilon^\mu \hat{k}^\nu)\;,
\end{equation}
where prime denotes derivative with respect to the phase $\xi$. Notice that contraction of this tensor with $\hat{k}_\mu$ is zero because of properties  \req{eq:properties}. Another quantity of interest seen in \req{eq:seconddynam} is
\begin{equation}\label{eq:dualtensor}
(s\cdot \partial)F^{*\mu\nu} = \frac{\mathcal{A}_0\omega^2}{c^2} f''(\xi) (\hat{k} \cdot s) \epsilon^{\mu\nu\alpha\beta}\hat{k}_\alpha\varepsilon_\beta \;,
\end{equation}
this time the contraction with both $\hat{k}_\mu$ or $\varepsilon_\mu$ is zero because of the anti-symmetry properties of $\epsilon^{\mu\nu\alpha\beta}$. 

We rewrite the dynamical Eqs.\;(\ref{eq:firstdynam},\;\ref{eq:seconddynam}) for the plane wave field using Eqs.\;(\ref{eq:emtensor},\;\ref{eq:dualtensor})
\begin{align}
\label{eq:firstplanewave}
\dot{u}^\mu &= - \frac{\mathcal{A}_0d\omega^2}{mc^2} f''(\xi) (\hat{k} \cdot s) \epsilon^{\mu\nu\alpha\beta}u_\nu \hat{k}_\alpha\varepsilon_\beta\;,\\
\dot{s}^\mu &= \omega d \mathcal{A}_0 f'(\xi) (\hat{k}^\mu \varepsilon \cdot s - \varepsilon^\mu \hat{k} \cdot s) - u^\mu (u \cdot F \cdot s) \frac{d}{c} \nonumber \\
&-\frac{\mathcal{A}_0 d \omega^2}{mc^2} f''(\xi) (\hat{k} \cdot s) \epsilon^{\mu\nu\alpha\beta}s_\nu \hat{k}_\alpha\varepsilon_\beta\label{eq:secondplanewave}\;.
\end{align}
In order to solve this system of equation we first look on the dot product of these equations with $\hat{k}_\mu$ and $\varepsilon_\mu$ which allows us to find a differential equation for $\hat{k} \cdot s(\tau)$ (Section \ref{sec:projections}). Then we can solve for the particle dynamics $u^\mu(\tau)$ (Section \ref{sec:particledynamics}) and invariant acceleration (Section \ref{sec:invacc}). Finally we will look on the motion in the laboratory frame (Section \ref{sec:labframe}). 

\subsection{Solutions for the spin projections}\label{sec:projections}
Contracting the first dynamical \req{eq:firstplanewave} with $k_\mu$ we  obtain a first integral of motion
\begin{equation}\label{eq:firstintegral}
\hat{k} \cdot \dot{u} = \frac{d}{d\tau} (\hat{k} \cdot u) = 0, \quad \Rightarrow \quad \hat{k} \cdot u = \hat{k} \cdot u(0)\;.
\end{equation}
This also allows us to find a relationship between the phase of the wave $\xi$ and proper time of the particle
\begin{equation}\label{eq:phase}
\frac{d\xi}{d\tau} = \frac{\omega}{c}\frac{d}{d\tau}(\hat{k} \cdot x) = \frac{\omega}{c}\hat{k} \cdot u(0), \; \Rightarrow \; \xi = \frac{\omega}{c}(\hat{k} \cdot u(0)) \tau + \xi_0\;.
\end{equation}

Repeating the same line of argument with $\varepsilon_\mu$ yields a second integral of motion
\begin{equation}\label{eq:secondintegral}
\varepsilon \cdot \dot{u} = \frac{d}{d\tau} (\varepsilon \cdot u) = 0, \quad \Rightarrow \quad \varepsilon \cdot u = \varepsilon \cdot u(0)\;.
\end{equation}

Now we consider contractions of the second dynamical \req{eq:secondplanewave}. Using the first integral of motion \req{eq:firstintegral} the contraction with $k_\mu$ reads
\begin{equation}\label{eq:kprod}
\hat{k} \cdot \dot{s} =  - (\hat{k} \cdot u(0)) (u \cdot F \cdot s)\frac{d}{c}\;.
\end{equation}
Using now the second integral of motion \req{eq:secondintegral} the contraction with $\varepsilon_\mu$ is
\begin{equation}\label{eq:epsilonprod}
\varepsilon \cdot \dot{s} =  \omega d A_0 f'(\xi) (\hat{k} \cdot s) - (\varepsilon \cdot u(0)) ( u \cdot F \cdot s)\frac{d}{c}\;.
\end{equation}
The scalar quantity $u \cdot F \cdot s$ can be evaluated using \req{eq:emtensor} and integrals of motion Eqs.\;(\ref{eq:firstintegral},\;\ref{eq:secondintegral})
\begin{equation}\label{eq:ufs}
u \cdot F \cdot s = \frac{\mathcal{A}_0\omega}{c} f'(\xi) W(\tau)\;,
\end{equation}
where
\begin{equation}\label{eq:wdef}
W(\tau) \equiv (\hat{k} \cdot u(0)) (\varepsilon \cdot s(\tau)) - (\varepsilon \cdot u(0)) (\hat{k} \cdot s(\tau))\;.
\end{equation}
This allows us to write \req{eq:kprod} as
\begin{equation}\label{eq:kdsw}
\hat{k} \cdot \dot{s} = - \mathcal{A}_0 \dot{f}(\xi(\tau)) W(\tau)\frac{d}{c}
\end{equation}
where we absorbed the $\omega (\hat{k} \cdot u(0))/c$ factor into the time derivative using differential of \req{eq:phase}. 

We now take another proper time derivative of this expression
\begin{equation}\label{eq:kdds}
\hat{k} \cdot \ddot{s} = -\mathcal{A}_0 \ddot{f}(\xi(\tau))W(\tau) \frac{d}{c} - \mathcal{A}_0 \dot{f}(\xi(\tau)) \dot{W}(\tau) \frac{d}{c}
\end{equation}
The term containing $\dot{W}(\tau)$ can be simplified by plugging both projections \req{eq:kprod} and \req{eq:epsilonprod} into the derivative of the definition $W(\tau)$ \req{eq:wdef} where the terms with $(u \cdot F \cdot s)$ cancel leaving us with
\begin{equation}
\dot{W}(\tau) = c d \mathcal{A}_0 \dot{f}(\xi(\tau)) (\hat{k} \cdot s)\;.
\end{equation}
Next we note that the first term on the RHS of \req{eq:kdds} can be expressed again in terms of $(\hat{k} \cdot \dot{s})$ using \req{eq:kdsw} leading us to the final dynamical equation for the spin projection
\begin{equation}
\label{Eq:SpinProjection}
\hat{k} \cdot \ddot{s} = \frac{\ddot{f}(\xi(\tau))}{\dot{f}(\xi(\tau))} (\hat{k} \cdot \dot{s}) - \mathcal{A}_0^2d^2 \dot{f}^2(\xi(\tau))(\hat{k} \cdot s)\;.
\end{equation}

Equation\;(\ref{Eq:SpinProjection})  is a second order linear differential equation for $\hat{k} \cdot s(\tau)$. The two general solutions are
\begin{equation}
\hat{k} \cdot s(\tau) = \exp\left(\pm i \mathcal{A}_0 d f(\xi(\tau)) \right)
\end{equation}
as can be verified by a direct substitution. The true physical solution can be found as their linear combination satisfying initial conditions which we will denote as
\begin{equation}
\hat{k} \cdot s(\tau = 0) \equiv \hat{k} \cdot s(0), \quad \hat{k} \cdot \dot{s}(0) = - \mathcal{A}_0 \dot{f}(\xi_0) W(0)\frac{d}{c}\;,
\end{equation}
where the initial condition for the derivative is given by \req{eq:kdsw}. After algebraic manipulations which heavily use trigonometric identities we obtain as our final result
\begin{equation}\label{eq:kssolution}
	\begin{aligned}
		\hat{k} \cdot s(\tau) = \hat{k} \cdot &s(0) \cos \left[\mathcal{A}_0 d (f(\xi(\tau)) - f(\xi_0))\right] \\
		&- \frac{W(0)}{c} \sin \left[\mathcal{A}_0 d (f(\xi(\tau)) - f(\xi_0))\right]\;.
	\end{aligned}
\end{equation}
It can be checked that this result satisfies the original dynamical problem and the associated initial conditions. Note that if we consider a situation when the initial state of the particle is long before the arrival of the plane wave pulse and final state long after it departed $f(\xi(\tau)) = f(\xi_0)$ and the projection $\hat{k} \cdot s(\tau)$ returns to its initial configuration $\hat{k} \cdot s(0)$. 

Now we  turn  to solving for the projections $\varepsilon \cdot s(\tau)$. Eliminating $u \cdot F \cdot s$ from Eqs.\;(\ref{eq:kprod},\;\ref{eq:epsilonprod}) yields
\begin{equation}
\varepsilon \cdot \dot{s} = \omega d\mathcal{A}_0 f'(\xi)(\hat{k} \cdot s) + \frac{\varepsilon \cdot u(0)}{\hat{k} \cdot u(0)} \hat{k} \cdot \dot{s}\;
\end{equation} 
and armed with the knowledge of solution for the $\hat{k} \cdot s(\tau)$ \req{eq:kssolution} we can integrate this equation imposing an initial condition $\varepsilon \cdot s(\tau = 0) = \varepsilon \cdot s(0)$ 
\begin{equation}\label{eq:epsilondotssolution}
	\begin{aligned}
		\varepsilon &\cdot s(\tau) = \varepsilon \cdot s(0) \cos \left[\mathcal{A}_0 d (f(\xi(\tau)) - f(\xi_0))\right]\\
		&+ \left(\frac{c \hat{k} \cdot s(0)}{\hat{k} \cdot u(0)} - \frac{W(0)}{c} \frac{\varepsilon \cdot u(0)}{\hat{k} \cdot u(0)} \right) \sin \left[\mathcal{A}_0 d (f(\xi(\tau)) - f(\xi_0))\right]\;.
	\end{aligned}
\end{equation}
Again, long after the passing of the pulse the projection $\varepsilon \cdot s(\tau)$ reinstates itself to the initial condition $\varepsilon \cdot s(0)$ long before the pulse's arrival.


\subsection{Solution for the 4-velocity}\label{sec:particledynamics}
Given the solutions for the projections of the 4-spin as a function of proper time we can solve for the particle motion. We start with the first dynamical \req{eq:firstplanewave}: we divide by $f''(\xi(\tau))\hat{k} \cdot s(\tau)$ and take another derivative with respect to proper time
\begin{equation}
\frac{d}{d\tau}\left(\frac{\dot{u}^\mu(\tau)}{f''(\xi) \hat{k} \cdot s(\tau)}  \right) =  - \frac{\mathcal{A}_0d\omega^2}{mc^2}\epsilon^{\mu\nu\alpha\beta}\dot{u}_\nu(\tau) \hat{k}_\alpha \varepsilon_\beta\;. 
\end{equation}
We can substitute for $\dot{u}_\nu$ on the right hand side back from the original dynamical equation \req{eq:firstplanewave} and contract the two anti-symmetric tensors while using integrals of motion Eqs.\,(\ref{eq:firstintegral}, \ref{eq:secondintegral}) as follows
\begin{align}
\epsilon^{\mu\nu\alpha\beta} &\epsilon_{\nu\rho\gamma\delta} \hat{k}_\alpha \varepsilon_\beta u^\rho(\tau) \hat{k}^\gamma  \varepsilon^\delta = \left| \begin{matrix}
\delta^\mu_\rho & \delta^\mu_\gamma & \delta^\mu_\delta \\
\delta^\alpha_\rho & \delta^\alpha_\gamma & \delta^\alpha_\delta \\
\delta^\beta_\rho & \delta^\beta_\gamma & \delta^\beta_\delta\\
\end{matrix} \right| \hat{k}_\alpha \varepsilon_\beta u^\rho(\tau) \hat{k}^\gamma  \varepsilon^\delta = \nonumber\\
&= \left| \begin{matrix}
u^\mu(\tau) & \hat{k}^\mu & \varepsilon^\mu\\
\hat{k} \cdot u(0) & 0 & 0 \\
\varepsilon \cdot u(0) & 0 & -1 
\end{matrix} \right| = (\hat{k} \cdot u(0))\hat{k}^\mu \;.
\end{align}
This gives us an expression
\begin{equation}\label{eq:interimstep}
\frac{d}{d\tau}\left(\frac{\dot{u}^\mu(\tau)}{f''(\xi) \hat{k} \cdot s(\tau)}  \right) = \frac{\mathcal{A}_0^2d^2\omega^4}{m^2c^4} (\hat{k} \cdot s(\tau)) f''(\xi)(\hat{k} \cdot u(0))\hat{k}^\mu\;,
\end{equation}
which can be formally integrated by introducing a unit-less function
\begin{equation}\label{eq:htau}
h(\tau) \equiv - \frac{\mathcal{A}_0d\omega^2}{mc^2}\int_{\tau_0 = 0}^\tau \hat{k} \cdot s(\widetilde{\tau}) f''(\xi(\widetilde{\tau}))d\widetilde{\tau}\;,
\end{equation}
this integral has to be computed for a specific laser pulse field. Let's integrate equation \req{eq:interimstep} twice using initial conditions 
\begin{align}
u^\mu(\tau_0 = 0) &\equiv u^\mu(0)\;, \\
\dot{u}^\mu(\tau_0 = 0) &= - \frac{\mathcal{A}_0d\omega^2}{mc^2} f''(\xi_0)(\hat{k} \cdot s(0)) \epsilon^{\mu\nu\alpha\beta}u_\nu(0)\hat{k}_\alpha \varepsilon_\beta\;,
\end{align}
 where the initial condition for the derivative is given by \req{eq:firstplanewave}. We obtain the final solution
\begin{equation}\label{eq:usolution}
\begin{aligned}
u^\mu(\tau) = u^\mu(0) &+ \frac{1}{2}h^2(\tau)\hat{k}^\mu (\hat{k} \cdot u(0)) +\\
	&+ h(\tau) \epsilon^{\mu\nu\alpha\beta}u_\nu(0)\hat{k}_\alpha \varepsilon_\beta\;.
\end{aligned}
\end{equation}
Finally, we also want to evaluate the expression 
\begin{equation}\label{eq:epsukeps}
\epsilon^{\mu\nu\alpha\beta}u_\nu \hat{k}_\alpha \epsilon_\beta = \epsilon^{\mu\nu\alpha\beta}u_\nu(0) \hat{k}_\alpha \epsilon_\beta + h(\tau) (\hat{k} \cdot u(0)) \hat{k}^\mu\;,
\end{equation}
where we used the solution \req{eq:usolution} and contraction identity for the anti-symmetric tensors. This equation will prove to be useful in the Section \ref{sec:labframe} when we evaluate the motion in the laboratory frame as it  captures  the motion in the plane perpendicular to the wave vector and the polarization direction. Three 4-vectors $k^\mu,\; \varepsilon^\mu$, and $\epsilon^{\mu\nu\alpha\beta}u_\nu(0) k_\alpha\varepsilon_\beta$ are mutually 4D-orthogonal and can be taken as a covariant basis of a 3D-subspace of the Minkowski space. 

\subsection{Invariant acceleration of the particle}\label{sec:invacc}

Another quantity of interest is the invariant acceleration which can be obtained by squaring \req{eq:firstplanewave} and contracting the anti-symmetric tensors
\begin{equation}\label{eq:invacc}
\dot{u}^2(\tau) = - \left(\frac{d\mathcal{A}_0\omega^2}{mc^2} \right)^2 f''(\xi)^2 (\hat{k} \cdot s(\tau))^2 (\hat{k} \cdot u(0))^2\;. 
\end{equation}
This expression is completely defined by the second derivative of the pulse function $f''(\xi)$ and by the derived solution for $\hat{k} \cdot s(\tau)$ \req{eq:kssolution}.

\subsection{Laboratory frame quantities}\label{sec:labframe}
\begin{figure}
	\begin{center}
		\includegraphics[width=.5\textwidth]{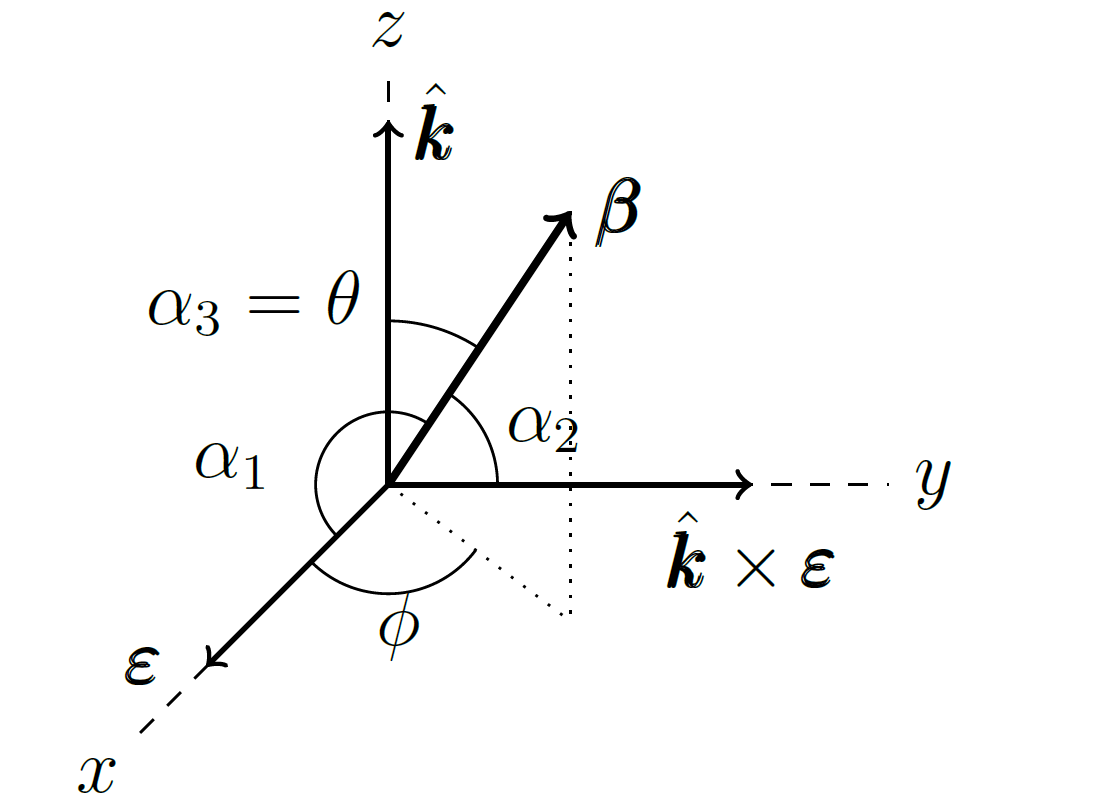}
		\caption{\label{fig:coordinates} Coordinates in the laboratory frame are chosen so that unit vector $\pmb{\varepsilon}$ points in the direction of $x$-axis, $\hat{\pmb{k}}\times\pmb{\varepsilon}$ in the direction of $y$-axis, and $\hat{\pmb{k}}$ in the direction of $z$-axis. $\alpha_1, \alpha_2$, and $\alpha_3$ are direction cosines with respect to these axes respectively.}
	\end{center}
\end{figure}
In the laboratory (lab) frame the relevant initial 4-vectors read
\begin{equation}
u^\mu(0) = \gamma_0 c (1, \pmb{\beta}_0), \quad \hat{k}^\mu = (1, \hat{\pmb{k}}), \quad \varepsilon^\mu = (0, \pmb{\varepsilon})\;.
\end{equation}
The initial spin 4-vector can be obtained by imposing a condition $u(0) \cdot s(0) = 0$ and Lorentz transformation of the instantaneous co-moving frame spin $(0,\pmb{s}_0)$
\begin{equation}
s^\mu(0) = (\pmb{\beta}_0 \cdot \pmb{s}_{0L}, \pmb{s}_{0L}), \quad \pmb{s}_{0L} = \pmb{s}_0 + \frac{\gamma_0 - 1}{\beta_0^2}(\pmb{\beta}_0 \cdot \pmb{s}_0) \pmb{\beta}_0\;.
\end{equation}

All initial conditions can be expressed in terms of the laboratory frame quantities as follows
\begin{align}
\hat{k} \cdot u(0) &= \gamma_0 c(1 - \hat{\pmb{k}} \cdot \pmb{\beta}_0)\;,\label{eq:labku0}\\
\varepsilon \cdot u(0) &= - \gamma_0 c \pmb{\varepsilon} \cdot \pmb{\beta}_0 \;,\\
\varepsilon \cdot s(0) &= - \pmb{\varepsilon} \cdot \pmb{s}_0 - (\gamma_0 - 1)(\hat{\pmb{\beta}}_0 \cdot \pmb{s}_0) (\pmb{\varepsilon} \cdot \hat{\pmb{\beta}}_0)\;,\\
\hat{k} \cdot s(0) &= \gamma_0 \pmb{\beta}_0 \cdot \pmb{s}_0 - \hat{\pmb{k}} \cdot \pmb{s}_0 - (\gamma_0 - 1)(\hat{\pmb{\beta}}_0 \cdot \pmb{s}_0)(\hat{\pmb{k}} \cdot \hat{\pmb{\beta}}_0)\label{eq:labks0}\;.
\end{align}
The first integral of motion \req{eq:firstintegral} can be expressed in the lab frame as
\begin{equation}\label{eq:labfirstint}
\gamma (1 - \hat{\pmb{k}} \cdot \pmb{\beta}) = \gamma_0 (1 - \hat{\pmb{k}} \cdot \pmb{\beta}_0)\;,
\end{equation}
and similarly for the second integral of motion \req{eq:secondintegral}
\begin{equation}\label{eq:labsecondint}
\gamma \pmb{\varepsilon} \cdot \pmb{\beta} = \gamma_0  \pmb{\varepsilon} \cdot \pmb{\beta_0} \;.
\end{equation}
Let's define 
\begin{equation}\label{eq:gammatau}
\gamma(\tau) \equiv \gamma_0 (1 + G(\tau))\;,
\end{equation}
where $G(\tau)$ is a measure of how big is the difference between the instantaneous $\gamma$-factor $\gamma(\tau)$ and initial $\gamma$-factor $\gamma_0$ relative to $\gamma_0$. Now the zeroth component of the 4-velocity solution \req{eq:usolution} tells us how is the $\gamma$-factor changing as a function of the particle's proper time. In terms of $G(\tau)$ we have
\begin{equation}\label{eq:gtau}
G(\tau) \equiv \frac{1}{2}h^2(\tau)(1 - \hat{\pmb{k}} \cdot \pmb{\beta}_0) + h(\tau) \pmb{\beta}_0 \cdot (\hat{\pmb{k}} \times \pmb{\varepsilon})\;.
\end{equation}
This expression also allows us to find the magnitude of velocity as a function of proper time
\begin{equation}\label{eq:betatau}
\beta^2(\tau) = 1 - \frac{1-\beta_0^2}{(1+G(\tau))^2}\;. 
\end{equation}
Taking a zeroth component of Eq.\, \req{eq:epsukeps} gives us 
\begin{equation}\label{eq:betadokkeps}
\gamma \pmb{\beta} \cdot (\hat{\pmb{k}} \times \pmb{\varepsilon}) = \gamma_0 \pmb{\beta}_0 \cdot (\hat{\pmb{k}} \times \pmb{\varepsilon}) + \gamma_0 h(\tau) (1 - \hat{\pmb{k}} \cdot \pmb{\beta}_0) \frac{\omega}{c}\;.
\end{equation}

We introduce three directional cosines which are projections of the $\pmb{\beta}$ vector on each of the coordinate axes in direction of unit vectors, $\pmb{\varepsilon}$, $\hat{\pmb{k}} \times \pmb{\varepsilon}$, and $\hat{\pmb{k}}$ shown in Fig.\;\ref{fig:coordinates}
\begin{equation}\label{eq:cosines}
\pmb{\varepsilon} \cdot \pmb{\beta} = \beta \cos\alpha_1, \quad (\hat{\pmb{k}} \times \pmb{\varepsilon}) \cdot \pmb{\beta} = \beta \cos \alpha_2, \quad \hat{\pmb{k}} \cdot \pmb{\beta} = \beta \cos\alpha_3\;. 
\end{equation}
Using expressions for $\gamma(\tau)$ \req{eq:gammatau} and $\beta(\tau)$ \req{eq:betatau} in the second integral of motion \req{eq:labsecondint} determines how the first directional cosine $\alpha_1(\tau)$ changes
\begin{equation}
\cos \alpha_1(\tau) = \frac{\beta_0 \cos\alpha_1(0)}{\sqrt{\beta_0^2 + G^2(\tau) + 2 G(\tau)}}\;.
\end{equation}
Similarly, substituting into \req{eq:betadokkeps} gives us an expression for $\alpha_2(\tau)$
\begin{equation}
\cos \alpha_2(\tau) = \frac{\beta_0 \cos \alpha_2(0) + h(\tau)(1 - \beta_0 \cos \alpha_3(0))\omega/c}{\sqrt{\beta_0^2 + G^2(\tau) + 2 G(\tau)}}\;.
\end{equation}
And finally, the third directional cosine $\alpha_3(\tau)$ can be obtained by substituting all the quantities into  first integral of motion \req{eq:labfirstint}
\begin{equation}
\cos \alpha_3(\tau) = \frac{G(\tau) + \beta_0 \cos\alpha_3(0)}{\sqrt{\beta_0^2 + G^2(\tau) + 2 G(\tau)}}\;.
\end{equation}

Now we can easily switch to the usual spherical angles $\theta(\tau), \phi(\tau)$ using formulas
\begin{equation}
\cos\theta(\tau) = \cos \alpha_3(\tau), \quad \tan \phi(\tau) = \frac{\cos \alpha_2(\tau)}{\cos \alpha_1(\tau)}\;,
\end{equation}
leading us to
\begin{equation}\label{eq:thetatau}
\cos \theta(\tau) = \frac{G(\tau) + \beta_0 \cos \theta_0}{\sqrt{\beta_0^2 + G^2(\tau) + 2 G(\tau)}}\;,
\end{equation}
and 
\begin{equation}
\tan \phi(\tau) = \tan \phi_0 + h(\tau) \frac{\omega}{c} \left[\frac{1 - \beta_0 \cos\theta_0}{\beta_0 \sin \theta_0 \cos\phi_0} \right]\;.
\end{equation}
The meaning of the conserved quantities $k \cdot u(0)$ and $\varepsilon \cdot u(0)$ or in the laboratory frame the equations \req{eq:labfirstint} and \req{eq:labsecondint} is that particle can lower its velocity while increasing the angle $\theta$ of its direction of motion with respect to $\hat{\pmb{k}}$ and vice versa. It also means that the geometry dictates a minimal velocity for the particle given the initial conditions. From Eqs.\,(\ref{eq:labfirstint},\ref{eq:gammatau},\ref{eq:cosines})
\begin{equation}
G(\tau) > \frac{\beta_0^2\sin^2\theta_0}{2(\beta_0\cos\theta_0-1)}\;.
\end{equation} 

\section{Ultra-relativistic neutrinos}\label{sec:neutrinos}
\begin{figure}
	\begin{center}
		\includegraphics[width=.5\textwidth]{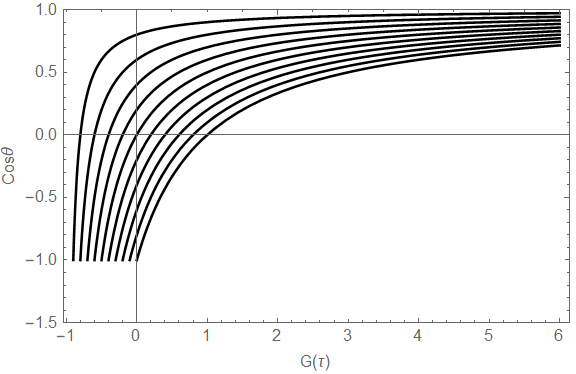}
		\caption{\label{fig:plot}Cosine of the angle between $\pmb{\beta}$ and $\hat{\pmb{k}}$ (Equation \req{eq:thetatau}) plotted as a function of $G(\tau)$ for different initial values $\cos\theta_0 \in [-1,1]$ which correspond to $G(\tau) = 0$. The ultra-relativistic limit $\beta_0 \approx 1$ is considered.}
	\end{center}
\end{figure}

The primary objective of this study is to show if Dirac neutrinos can be deflected in their path by intense laser fields. This would mean that the neutrino beams that are focused on detectors far away for the purpose of study of neutrino oscillations would experience variation in the event count as a function of applied pulsed laser field, where neutrino pulses and laser pulses are synchronized. We will estimate the magnitude of the effect as a function of both laser and neutrino properties.

We rewrite the amplitude of the laser field $\mathcal{A}_0$ using the dimensionless normalized amplitude $a_0$ 
\begin{equation}
\mathcal{A}_0 = \frac{m_e c}{e}a_0\;.
\end{equation}
The conversion between the magnetic moment and elementary dipole charge of particle $d$ reads
\begin{equation}
d = \frac{e}{mc}\mu_\nu[\mu_B]\;,
\end{equation}
where $\mu_\nu$ is in units of Bohr magnetons. This makes the product $A_0 d$ appearing in our equations
\begin{equation}\label{eq:estimate}
A_0 d = a_0 \mu_\nu[\mu_B] \approx 10^{-9} - 10^{-18}\;,
\end{equation}
for the state of the art laser systems with $a_0 \sim 10^2$ and whole range of possible magnetic moments of the Dirac neutrinos Eqs.\,(\ref{eq:lower},\ref{eq:upper}). Such laser system is currently under construction in ELI Beamlines, Prague and the dimensionless normalized amplitude for this laser corresponds to the power of 10 PW \cite{rus2015eli} with an intensity $10^{24}$ W/cm$^2$. Since the product $A_0 d$ is so small the arguments of the trigonometric functions in the solutions Eqs.\;(\ref{eq:kssolution},\;\ref{eq:epsilondotssolution}) are negligible and the solutions reduce to
\begin{equation}\label{eq:ultrarel}
\hat{k} \cdot s(\tau) \approx \hat{k} \cdot s(0), \quad \varepsilon \cdot s(\tau) \approx \varepsilon \cdot s(0)\;.
\end{equation}
In other words there is no precession in these directions. For ultra-relativistic neutrinos $\beta_0 \approx 1$ Fig.\;\ref{fig:plot} shows that if the laser field is increasing the velocity of the neutrinos in the beam for $G(\tau) > 0$, the neutrinos have tendency to focus - cosine of the angle between the velocity of neutrinos and wave vector of the laser light \req{eq:thetatau} approaches one as $\gamma(\tau)$ increases.

As we showed in our previous work \cite{formanek2017strong} the square root of the invariant acceleration \req{eq:invacc} is greatly increased by the initial gamma factor of the neutrino $E_\nu/m_\nu$ squared. Now that we have a solution for the particle motion we can estimate how the gamma factor \req{eq:gammatau} and the angle \req{eq:thetatau} of the neutrino are changing with the proper time.  

\subsection{Estimate of change for velocity and angle}
Both the angle \req{eq:thetatau} and the gamma factor \req{eq:gammatau} depend only on the value of the function $G(\tau)$. This function \req{eq:gtau} can be evaluated by explicitly integrating $h(\tau)$ \req{eq:htau} with a specific laser pulse oscillations and profile. In the ultra-relativistic limit \req{eq:ultrarel} this integral can be estimated as
\begin{equation}
h(\tau) \approx -\frac{\mathcal{A}_0 d \omega}{mc} \frac{\hat{k} \cdot s(0)}{\hat{k} \cdot u(0)} f'(\xi(\tau))\;,
\end{equation}
where we assumed that we started counting the proper time long before the pulse arrived when $f'(\xi_0) = 0$. From the lab frame quantities Eqs.\,(\ref{eq:labku0}, \ref{eq:labks0}) the fraction 
\begin{equation}
\frac{\hat{k} \cdot s(0)}{\hat{k} \cdot u(0)} \approx \frac{\pmb{\beta}_0 \cdot \pmb{s}_0}{c} \approx \frac{\hbar}{c}\;,
\end{equation} 
which does not depend on the initial gamma factor. Using the estimate \req{eq:estimate} the expression for $h(\tau)$ is
\begin{multline}\label{eq:htau}
h(\tau) \approx - \frac{a_0 \mu_\nu[\mu_B]E_\gamma[eV]}{m_\nu[eV]}f'(\xi(\tau))\\ \approx - (10^{-8} - 10^{-17}) f'(\xi(\tau))\;,
\end{multline}
where $E_\gamma$ is the energy of the laser photons and the numerical value was calculated for $E_\gamma = 1$ eV because ELI Beamlines will operate in the visible range. Our result \req{eq:htau} shows that the effect depends on the product of $a_0$ with $E_\gamma$ which is proportional to the square root of laser intensity. There are lasers with much higher electron energy (like free electron lasers XFEL in Hamburg \cite{altarelli2007european} or LCLS-II under construction in Stanford \cite{Galayda:2014qka}), but their intensity is lower for coherent photons from a given energy band. Moreover, lasers with higher photon energy have shorter wavelength which would invalidate our condition of classical limit \req{eq:classlim} for some sources (For example neutrinos from beta decay have energy $\sim$ 1 MeV which corresponds to $\lambda_\nu \sim 10^{-12}$ m and 10 keV photon has wavelength $\sim 10^{-10}$ m). Mass of neutrino was taken as $m_\nu = 0.2$ eV. This is for neutrinos even at the best case scenario a very small number because unlike acceleration this expression is not sensitive to the initial gamma factor $\gamma_0$. Looking at the equation for $G(\tau)$ - equation \req{eq:gtau} the square of $h(\tau)$ is completely negligible so that if we keep only the linear term in $h(\tau)$ we get
\begin{equation}
G(\tau) \approx (10^{-8} - 10^{-17}) f'(\xi(\tau))\;,
\end{equation} 
which means that unless we can prepare laser with very high derivative $f'(\xi)$ in which our approximation \req{eq:ultrarel} is no longer valid, the changes of the gamma factor and angle will be minuscule.
 
Note that this ultra-relativistic limit is also not valid for neutrons because the product $A_0 d$ is not negligible anymore and the projection $k \cdot s(\tau)$ is no longer constant  - the spin precesses - and has to be taken into account in the integral \ref{eq:htau}.  We wish to return to the neutron dynamics in the laser field in the future.  

\section{Discussion and conclusions}
In this paper we formulated and explored a covariant classical neutral particle dynamics in external EM fields. The neutral particle interacts with the fields through its magnetic moment and in the co-moving frame feels a Stern-Gerlach force \req{eq:force} and torque \req{eq:torque}. Our covariant formulation clarifies that apart from the usual torque term $\pmb{\mu}\times \pmb{B}$ we also have a term proportional to the gradient of the electric field in the direction of the spin which is necessary to satisfy the constraint of spin and 4-velocity orthogonality \req{eq:constraint1} while keeping the Stern-Gerlach force intact. 

In the proposed formulation we restricted ourselves to the natural form of the spin dynamics, however we cannot exclude that other terms orthogonal to $u^\mu$ can be added to the dynamical equation for spin. Therefore in the section \ref{sec:nonuni} we discuss the non-uniqueness of the torque equation and possibility of adding another term which would change the Thomas precession coefficient. While we continue our search for theoretical rationale that would uniquely define the torque equation, we note that a precession experiment with neutral particles with nonzero magnetic moment can determine if such modification is present.

We presented an analytical solution of our dynamical equations in the external linearly polarized electromagnetic plane wave field. We showed that the projections of particle 4-velocity on the wave and polarization 4-vectors are constants of the motion Eqs.\,(\ref{eq:firstintegral},\ref{eq:secondintegral}). We formulated a differential equation for the projection of the 4-spin $k \cdot s(\tau)$ and found its solution \req{eq:kssolution}. Finally, we solved for the 4-velocity of the particle \req{eq:usolution}.

Our results are obtained in the classical dynamics framework. Several decades ago Skobelev \cite{skobelev1991interaction} considered quantum field theory formulation of processes in the presence of the magnetic / electrical dipole of neutrino. For the state of the art fields of that time period the effect was not measurable. However, if we extrapolate the progress in laser technology made since this work appeared, we remain optimistic that experiments studying this effect will become possible. While quantum approach may provide additional motivation for selection of classical dynamical equations, considering the short de Broglie wavelength of ultra-relativistic particles, classical dynamics may suff how these results can be used in neutron and neutrino beam control, allowing in the case of neutrinos to obtain information about their properties.

\bibliography{references}{}
\bibliographystyle{ieeetr}
\end{document}